# Generating topological non-diffracting beams using high quality factor nonlocal metasurfaces

**Authors**

Dongha Kim[1,2], Charles Pelzman[3], Cheng Guo[1,4], Olivia Y. Long[1], Shanhui Fan[1*], and Sang Yeon Cho[3*]

**Affiliation**

[1]Ginzton Laboratory, Stanford University, Stanford, CA 94304, United States

[2]Department of Physics, Korea University, Seoul, 02483, Republic of Korea

[3]DEVCOM U.S. Army Research Laboratory, Adelphi, MD 20783, United States

[4]Department of Electrical and Computer Engineering and Microelectronics Research Center, University of Texas at Austin, Austin, TX 78712, United States

Email: shanhui@stanford.edu, sang.y.cho.civ@army.mil

**Abstract**

Non-diffracting optical beams are essential tools in photonics, offering robust light transport, super-resolution imaging, and spatiotemporal control. While nonlocal metasurfaces have been proposed for structured light generation due to their broad angular spectral dispersion and topological characteristics, experimental generation of non-diffracting beam with nonlocal metasurfaces have not been previously demonstrated. Here, we experimentally realize vortex Bessel beams using a nonlocal metasurface and uncover a link between non-diffracting beam generation and the band curvature of the photonic bands. Depending on the sign of band curvature, the beams exhibit spatial asymmetry in non-diffraction, emerging either in front of or behind the metasurface. This asymmetry arises from a radial phase gradient in momentum space, which induces effective space compression or expansion. Furthermore, we demonstrate wavelength-dependent tunability of the beam diameter and propagation distance, and show an order-of-magnitude enhancement in propagation distance compared to conventional Laguerre–Gaussian modes. These results establish nonlocal metasurfaces as a powerful platform for compact, tunable, and spatiotemporally controlled non-diffracting light.

# Introduction

Conventional optical beams diffract under propagation [1]. The suppression of diffraction is of long-standing interest in optics, as it enables robust, long-distance transmission of electromagnetic waves across various media [2-6]. For this purpose, there have been significant works on Bessel beams, which maintain its intensity distribution and shape during propagation [7-13]. One may in addition achieve a non-zero orbital angular momentum on a Bessel beam, creating a vortex Bessel beam that is diffraction-free and has non-trivial topological structures [14,15].

Bessel beams can be generated with conventional optical components such as annular apertures with lenses [8], axicons [9,10], spatial light modulators [11,12], and diffraction gratings [13]. In recent years, there have also been significant developments in using local metasurfaces as a more compact mechanism to generate Bessel beams [16-19]. A local metasurface consists of a space-dependent arrangement of meta-atoms. The shape and/or the orientation of the meta-atoms can be appropriately chosen to create a space-dependent complex transmission/reflection coefficients, so that an incident wave, upon interacting with the metasurface, can generate a Bessel beam. With this scheme, vortex Bessel beams can also be generated with suitable choices of the arrangements of the meta-atoms.

In contrast with the use of local metasurface, in this work, we experimentally demonstrate the use of nonlocal metasurface to generate Bessel and Bessel vortex beams (Figure 1a). A nonlocal metasurface consists of a periodic arrangement of scattering elements. The elements and their arrangements are specifically designed to control the wavevector dependency of the transport coefficients [20-27]. Here we show that a suitable design of the nonlocal metasurface leads to the generation of Bessel beams. Compared with local metasurfaces, since the nonlocal metasurface is periodic, there is no need for alignment in the relative position between the incident beam and the center of the metasurface. Also, the structure and the design of the nonlocal metasurface is simpler.

The concept of using nonlocal metasurface to generate Bessel beam was proposed theoretically in Ref. [28]. However, Ref. [28] did not provide an experimental demonstration of Bessel beam generation. The experiments in Ref. [28] demonstrated optical vortex beam generation for light transmitted through the metasurface. A key feature of the behaviors of nonlocal metasurface is the presence of guided resonances. As we will discuss in more detail below, in the nonlocal metasurfaces as considered in both Ref. [28] and this work, there is significant transmission in wavevector components away from the guided resonances. As a result, the transmitted light usually does not form a Bessel beam. In contrast with Ref. [28], in this paper we focus on reflection. With suitable design, the reflected light consists entirely of contributions from the guided resonances, which facilitates Bessel beam generation.

## Results

**Theoretical background: circular intensity distribution in nonlocal metasurface**

In the wavevector space, an ideal Bessel beam has an amplitude distribution $E(k_t)$ of the form

$$E(k_t, \phi_{kt}) \propto E(\phi_{kt}) \delta(k_t - k_{t,0}) \tag{1}$$

where $k_t$ is the magnitude of wavevector perpendicular to the beam propagation direction and $\phi_{kt}$ is the azimuthal angle in the wavevector space. Such ideal Bessel beam can propagate over arbitrary propagation distance without any diffraction, but is not physically realistic as it contains infinite power. In practical Bessel beam, the Dirac delta function in Eq. (1) is replaced by a function $f(k_t)$ that is narrowly peaked near $k_{t,0}$ but has a finite linewidth. Such practical Bessel beam has finite power, and can propagate over substantial distance before diffraction occurs [29,30].

A Bessel beam can have non-zero topological textures in its field distributions on the plane perpendicular to the propagation direction. In this paper, we will specifically consider a Bessel beam with optical vortices of the form

$$E(\phi_{kt}) \propto e^{il\phi_{kt}} \quad (2)$$

where $l$, an integer, is the winding number of optical vortex. Such an optical vortex in momentum space corresponds to a vortex beam in real space with the same winding number $l$.

In order to generate the Bessel beam as described in Eqs. (1) and (2), we consider the reflection when a circularly polarized Gaussian beam is normally incident upon a suitably designed nonlocal metasurface as shown in Figure 1a. In our design, we choose a nonlocal metasurface with a $C_{6v}$ symmetry, consisting of a triangular lattice of airholes introduced into a high-index dielectric thin film. As a consequence of the $C_{6v}$ symmetry, near the Γ point with the magnitude of the transverse wavevector $k_t = 0$, the constant frequency contour of the guided resonance in the wavevector space becomes circular [31].

In general, a nonlocal metasurface supports guided resonance [32]. For this system with the $C_{6v}$ symmetry, near the guided resonance, the reflection coefficient is dependent upon frequency ($\omega$) and the magnitude of the transverse wavevector ($k_t$) in the form of:

$$r(k_t, \omega) = r_d + \frac{\gamma}{-i(\omega - \omega(k_t)) + \gamma}$$

where $r_d$ is the background reflection and $\gamma$ is the damping rate. The $\omega(k_t) \sim \omega_0 + \alpha k_t^2$ is the dispersion relation near Γ point with $k_t = 0$, where $\omega_0$ is the resonance frequency at Γ point and $\alpha = \frac{1}{2}\frac{\partial^2 \omega(k_t)}{\partial k_t^2}$ is the band curvature.

Moreover, with a suitable choice of the thickness of the film, the background reflection coefficient $r_d$ vanishes. Assuming that $\gamma$ is sufficiently small, for an incident beam with an operating frequency $\omega_b$, the reflected light in cylindrical coordinate $(r, \theta, z)$ has the approximate form:

$$E(r, \theta, z) = AJ_l(k_t^b r) \exp(ik_z z) \exp(il\theta)$$

where $A$ is the amplitude of electric field, $k_t^b = \sqrt{(\omega_b - \omega_0)/\alpha}$, and $J_l(k_t^b r)$ is the $l$-th order Bessel function.

The helical phase distribution in the reflected light arises from the symmetry of guided resonance at the Γ point. As a consequence of the $C_{6v}$ symmetry group, the irreducible representations of A1, A2 (B1, B2) modes can give rise to topological polarization texture in wavevector space, carrying topological charge of $q = \pm 1$ ($\pm 2$) [31, 33]. When circularly-polarized incident light interacts with these textures, the reflected light in the opposite polarization acquires a geometric phase, giving rise to an optical vortex with winding number $l = -2q$ [28]. Consequently, the reflected field can generate vortex Bessel beams with $l = \pm 2, \pm 4$, as illustrated in Figure 1b.

The use of such guided resonance band to produce a Bessel beam has been theoretically proposed in Ref. [28]. In the experiments, Ref. [28] measured transmitted light through nonlocal metasurface for the optical vortex generation. In order to generate an optical vortex, one only needs to ensure that the phase has a dependency of $e^{il\theta}$ on the azimuthal angle, without the need to control the intensity dependence as a function of radius $r$. In contrast, to create a Bessel beam one also needs to control the radial distribution of the intensity. In particular, the form of Eq. (1) indicates that the Fourier components of the Bessel beam should be non-zero only in the vicinity of $k_t = k_t^b$. Such a requirement is difficult to achieve in the transmission of the nonlocal metasurface since the background transmission in general is non-zero. Therefore, in our study, we measured reflected light and moreover specifically configure our nonlocal metasurface to achieve zero background reflection and thus facilitate Bessel beam generation. Furthermore, we observed that the measured reflected light reveals unexpected spatial asymmetry in its propagation profile depending on the operating wavelength as shown in Figure 1c. The curvature of photonic bands can induce space compression or expansion effect on the vortex Bessel beam, which will be discussed in Figures 3 and 4.

**Photonic band structure characterization**

We designed and fabricated a triangular-lattice silicon photonic crystal to generate the vortex Bessel beams. The structure consists of a multilayer stack (SiO$_2$ substrate / a-Si 170 nm / SiO$_2$ 10 nm / a-Si 90 nm), where a triangular array of holes is patterned in the 90-nm-thick a-Si layer. The periodicity and hole radius are 885 nm and 375 nm, respectively. The 10-nm SiO$_2$ layer serves as a stop layer for the dry etching process. Details of the fabrication procedure are provided in the Methods section. Using rigorous coupled-wave analysis [34], we calculated the reflectance distribution in ($k_t$, $\lambda$) space near the Γ point for the wavelength range 1300–1600 nm. The calculations reveal four distinct quadratic photonic bands with positive and negative curvatures (Supplementary Note S1).

We experimentally characterized the band structure of the fabricated triangular photonic crystal using back focal plane imaging. Figure 2a presents the measured reflectance spectra in ($k_t$, $\lambda$) space, showing four distinct photonic bands with different curvatures. We label these bands as the 1st band (1400–1425 nm), 2nd band (1500–1512 nm), 3rd band (1512–1550 nm), and 4th band (1537–1580 nm), respectively, according to their spectral positions. A three-dimensional visualization of the photonic bands in ($k_x$, $k_y$, $\lambda$) space is shown in Figure 2b.

In Figure 2c, we plot the intensity distribution $R(k_x, k_y)$ and the corresponding interferograms of the reflected light in momentum space. We selected representative wavelengths for each band: 1420 nm (1st band), 1506 nm (2nd band), 1515 nm (3rd band), and 1550 nm (4th band). In all cases, a ring-shaped dispersion with a narrow transverse wavevector linewidth was observed. The interferograms reveal spiral fringe patterns, confirming the presence of phase singularities. Based on the number and handedness of the spirals, the winding numbers were determined as +2 (1420 nm), −4 (1506 nm), +4 (1515 nm), and +4 (1550 nm), respectively.

To further quantify the dispersion, we extracted the transverse wavevector ($k_t$) and transverse linewidth ($\delta k_t$) distributions from the $R(k_x, k_y)$ maps (Figures 3d and 3e). The $k_t$ distributions along the K and M directions match well at low $k_t$, confirming the formation of an isotropic ring-shaped dispersion in ($k_x$, $k_y$) space. The hexagonal dispersion at larger $k_t$ also supports non-diffracting propagation, as discussed in Supplementary Note S2. The $\delta k_t$ values are smaller than $k_t$ by about two orders of magnitude (~$10^{-2}$), satisfying the non-diffraction condition. Finally, we extracted the quality factors ($Q$) of the four photonic bands from the reflectance spectra, as shown in Figure 2f. The $Q$ factors are considerably high (~$10^3$ to ~$10^4$) across all wavelength ranges, consistent with the observed narrow $\delta k_t$. Moreover, the diverging behavior of $Q$ near the $\Gamma$ point provides clear evidence of symmetry-protected bound states in the continuum (BICs) [35,36].

**Vortex Bessel beam generation**

We characterized the propagation profiles of the reflected light from the nonlocal metasurface by translating the infrared camera along the optical axis to record the transverse intensity distribution $I(x, y)$ at various propagation distances $z$. Figure 3a presents both the measured real-space intensity profiles and interferograms for four representative wavelengths—1410 nm, 1506 nm, 1512 nm, and 1550 nm—corresponding to the four distinct photonic bands. The intensity profiles exhibit donut-shaped patterns, characteristic of optical vortex beams. Spiral fringes observed in the interferograms confirm the presence of phase singularities, with winding numbers consistent with those extracted from the momentum-space interferograms in Figure 2c.

The measured real-space propagation profiles are shown in Figure 3b. The reflected beams maintain their transverse shape over extended propagation distances, demonstrating non-diffracting behavior. Notably, non-diffraction occurs asymmetrically: either on the −$z$ side (for wavelengths 1420 nm and 1506 nm) or on the +$z$ side (for wavelengths 1515 nm and 1550 nm). Here $z = 0$ corresponds to the position where the physical location of the metasurface is mapped to using the imaging optics. This spatial asymmetry has not been previously reported with other schemes of Bessel beam generation. Here we refer to this effect as the effect of asymmetric non-diffraction. The asymmetric behavior is further illustrated in the longitudinal intensity distributions $I(z)$, shown in Figure 3c, which are extracted along the propagation axis through the beam center (highlighted in Figure 4b). Figure 3d is the summary of winding number and spatial asymmetry of non-diffraction, depending on the operating wavelength.

**Asymmetric non-diffraction induced by space compression and space expansion**

The key distinction between the non-diffraction observed on the +$z$ and −$z$ sides lies in the band curvature ($\alpha$) of the photonic bands. The 1st and 2nd bands possess positive curvature ($\alpha > 0$), while the 3rd and 4th bands exhibit negative curvature ($\alpha < 0$). To investigate the effect of band curvature, we analyzed the amplitude $|r|$ and phase $\phi_r$ distributions of the reflection coefficient $r(k_x, k_y)$ of the guided resonance in momentum space. Figure 4a shows simulated distributions of $r(k_x, k_y)$ for resonances with opposite band curvatures: $\alpha = +2 \times 10^3$ rad·m²/s (red box) and $\alpha = -2 \times 10^3$ rad·m²/s (blue box), with $\omega_0$ corresponding to 1555 nm and 1545 nm, respectively. The quality factor $Q = 10^4$ was assumed for both cases. While the amplitude distributions $|r(k_x, k_y)|$ are identical, the phase distributions $\phi_r(k_x, k_y)$ show opposite radial gradients of $\partial \phi_r / \partial k_t$: negative for $\alpha > 0$ and positive for $\alpha < 0$.

Using these reflection coefficient distributions, we calculated the longitudinal intensity profiles $I(z)$ of the generated beams, based on [37,38]:

$$I(z) = \left| \iint_{-\infty}^{+\infty} r(k_x, k_y) e^{i(k_x x + k_y y + k_z z)} dk_x dk_y \right|^2$$

The results, shown in Figure 4c, reveal that non-diffraction emerges in the $-z$ direction for $\alpha > 0$ (red curve) and in the $+z$ direction for $\alpha < 0$ (blue curve). This confirms that the asymmetric non-diffraction originates from the radial phase response in momentum space, dictated by the band curvature. In contrast, when the phase response is uniform (green curve), the propagation profile is symmetric about $z = 0$, as expected for conventional non-diffracting beams.

The phenomenon can be further understood through the concepts of space compression and expansion induced by nonlocal metasurfaces. A quadratic phase distribution in momentum space is equivalent to the compression or expansion of free-space propagation distance [25–27]. To examine this effect, we analyzed the $\phi_r(k_t)$ distributions along the $k_t$-axis for different quality factors $Q = 10^3, 10^4$, with the opposite band curvature $\alpha = \pm 2 \times 10^3$ m$^2$/s (Figure 4c). The guided resonance parameters are identical to the Figure 4a,b, with the damping rate $\gamma$ determined from the quality factors $Q$. Near the amplitude maximum, $\phi_r(k_t)$ can be fitted by a quadratic function $\beta k_t^2$ (yellow line), where the magnitude of curvature parameter $\beta$ increases from $1.16 \times 10^{-9}$ m$^2$ to $1.20 \times 10^{-8}$ m$^2$, as $Q$ increases. Since the amplitude $|r(k_x, k_y)|$ is strongly peaked near the wavevectors of the guided resonance, the resulting beam propagation is dominantly governed by space expansion for $\alpha < 0$, or space compression for $\alpha > 0$. The effective propagation shift $d_{\text{eff}}$ induced by the phase curvature can be expressed as $d_{eff} = 4\pi\beta/\lambda$ where $\lambda$ is the operating wavelength. The calculated $d_{\text{eff}}$ are $-9.39$ mm ($Q = 10^3$, $\alpha < 0$), $-97.2$ mm ($Q = 10^4$, $\alpha < 0$), $97.2$ mm ($Q = 10^4$, $\alpha > 0$), indicating that the beam center is shifted to $z = -d_{\text{eff}}$ relative to $z = 0$.

Figure 4d compares the longitudinal intensity profiles $I(z)$ calculated from three different phase distributions: uniform (green), quadratic fit (yellow), and the full guided resonance phase response (blue), using the same amplitude $|r(k_x, k_y)|$ at a given $Q$. The uniform phase case results in symmetric propagation centered at $z = 0$, while the quadratic phase induces a shift of the beam center at $z = -d_{\text{eff}}$. The guided resonance phase response lies between the uniform and the quadratic phase cases, generating a non-diffracting beam with spatial asymmetry. Importantly, reversing the sign of $\alpha$ reverses the radial phase gradient and, consequently, the direction of beam shift: space compression for $\alpha > 0$ results in non-diffraction toward the $-z$ side. Additionally, higher $Q$ values lead to longer propagation distances, owing to the combined effects of reduced transverse wavevector linewidth and enhanced space compression/expansion.

**Wavelength-dependent tunability of vortex Bessel beam propagation**

Finally, we demonstrate that the nonlocal metasurface enables wavelength-dependent control over the propagation characteristics of vortex Bessel beams. Figure 5a shows the measured real-space propagation profiles of beams generated from the 4th photonic band at various operating wavelengths.

As the wavelength increases from 1560 nm to 1580 nm, the beam diameter varies from 94 μm to 66 μm, and the propagation distance varies from 22.8 mm to 13.4 mm. Figure 5b presents the extracted beam diameter $D$ and propagation distance $L$ as functions of the operating wavelength. In the entire wavelength range, the beam diameter varies from 66 μm to 329 μm, and the propagation distance varies from 11.4 mm to 107.0 mm. Both parameters exhibit strong tunability near the $\Gamma$ point of the photonic band. This behavior is expected, as the beam diameter $D$ and the propagation distance $L$ scale inversely with the numerical aperture and the linewidth of the longitudinal wavevector, respectively. To better illustrate this performance, we plotted the $D$-$L$ relation normalized by the operating wavelength $\lambda$ in Figure 5c. The resulting $D/\lambda$ and $L/\lambda$ values for our metasurface beams significantly exceed those of conventional Laguerre–Gaussian (LG) beams with the same winding number $l$. At a fixed transverse size $D/\lambda$, the metasurface-generated beams achieve a propagation distance $L/\lambda$ that is enhanced by an order of magnitude compared to LG beams.

To understand the origin of this enhancement, we further analyzed the dependence of $D$ and $L$ on the quality factor $Q$ of the guided resonance using the same simulation parameters as in Figure 4b. As shown in Figure 5d, the beam diameter $D$ (green) increases only slightly, by approximately 5%, despite a two-orders-of-magnitude increase of $Q$. In contrast, the propagation distance $L$ (blue) increases linearly with $Q$ from 15.8 mm to 803 mm, due to the narrowing of the longitudinal wavevector linewidth [37,38]. These results demonstrate that by optimizing the quality factor of the guided resonance, the propagation range of the vortex Bessel beam can be significantly extended without substantial growth of the beam diameter, enabling compact, long-range beam shaping in metasurface-based optical systems.

**Discussion**

In this work, we experimentally realize optical vortex Bessel beams using a high-Q nonlocal metasurface. We uncover an unexpected propagation behavior, which we term asymmetric non-diffraction, in which the intensity of vortex Bessel beam emerges selectively either in front of or behind the location of the metasurface. This asymmetry originates from the phase response of guided resonances: a radial phase gradient induces effective spatial compression or expansion of the vortex Bessel beam, with the sign set by the photonic band curvature. We further show a large, resonance-enabled tunability of the beam diameter and propagation distance via operating wavelength detuning. These dispersion-driven controls, governed by the resonance quality factor and phase dispersion, establish a compact, planar platform for programmable structured-light generation.

We envision that this platform will open new directions for the generation and dynamic control of spatiotemporally structured, non-diffracting light in nanophotonic environments. The nonlocal metasurface architecture offers a route toward miniaturizing complex bulk optics traditionally required to create exotic wavepackets, such as space-time wavepackets [37-39], toroidal pulses with embedded non-diffraction [40,41], and topological textures in space-time domains [42,43]. In future implementations, incorporating optically active materials [44-46] may allow for real-time tuning of the photonic band structure—enabling dynamic control of band curvature, resonance frequency, and quality factor, and thereby offering high-speed modulation of the beam's propagation distance, non-diffraction direction, and topological charge. In practical applications, compactness and alignment-free characteristics of nonlocal metasurface can widen the utilization of Bessel beams in bio-imaging [48], non-linear phase matching [49], and free-space communications [50]. These advances could enable new classes of ultrafast beam shaping devices and topological light sources with unprecedented spatiotemporal control.

## Methods

### Sample Fabrication

The metasurface was fabricated on a 500-um-thick fused-silica substrate. An amorphous silicon and silicon dioxide layers were deposited on a single side of the substrate with thicknesses of 170 nm/a-Si, 10 nm/$SiO_2$ and 90nm/a-Si using plasma-enhanced chemical vapor deposition. An ~350-nm-thick layer of electron-beam photoresist, ZEP520-A, was spin-coated on to the deposited layered side. A 20nm thick Au layer was thermally deposited on the electron-beam resist to prevent electrostatic charging during lithography. The nanostructure was patterned in the electron-beam resist through an electron-beam-lithography (EBL) system. The periodicity and radius of the triangular lattice are 885 nm and 375 nm, respectively. After the EBL, the Au layer was removed and the electron-beam photoresist was developed. A hard mask of 20 nm/$Al_2O_3$ was deposited using an electron-beam, and the photoresist was then removed. The top amorphous silicon layer was etched with a reactive-ion-etch process removing 90nm. The hard mask of $Al_2O_3$ was removed in a bath of AZ300MIF solution.

### Optical characterization

The photonic band structure of nonlocal metasurfaces is characterized by back focal plane imaging. Tunable laser (Santec TSL-570, TSL-770) illuminates the nonlocal metasurface with the right circularly polarized incidence, covering the operation wavelength range of 1400-1425 nm and 1500-1600 nm. The momentum space images of reflected light are collected by imaging the back focal plane of the objective lens using IR camera (Goldeye G/CL-030). The polarization state of the reflected light is analyzed into left circular polarization state. The propagation profiles of reflected light are collected by imaging the front focal plane of the objective lens, but scanning the longitudinal position of IR camera using motorized translation stage (Thorlabs LTS150C). The interferometric images are collected for both real-space and momentum space by introducing reference mirror as Michelson-type configuration. The detailed setup configuration is illustrated and discussed in Supplementary S4.

## Data Availability

The data that support the findings of this study are available from the corresponding authors upon reasonable request.

## Acknowledgement

D.K. acknowledges the support of the National Research Foundation of Korea (NRF, RS-2023-00240304) and Korea University Grant. C.G. acknowledges the support of the Jack Kilby/Texas Instruments Endowed Faculty Fellowship. S.F. acknowledges support from the U. S. Army Research Office (Grant No. W911NF-24-2-0170), and from a MURI project from the U. S. Air Force Office of Scientific Research (Grant No. FA9550-21-1-0312). S. Y. C. acknowledges support from the ARD–ARO joint program (W911NF2420170).

## Author Contribution

D. K. and S. Y. C. conceived the idea. C. P. fabricated triangular Si photonic crystal. D. K. conducted optical characterizations and numerical simulations. D. K., S. F., S. Y. C., wrote the manuscript and reflected the response from all authors, including C. G. and O. Y. L.

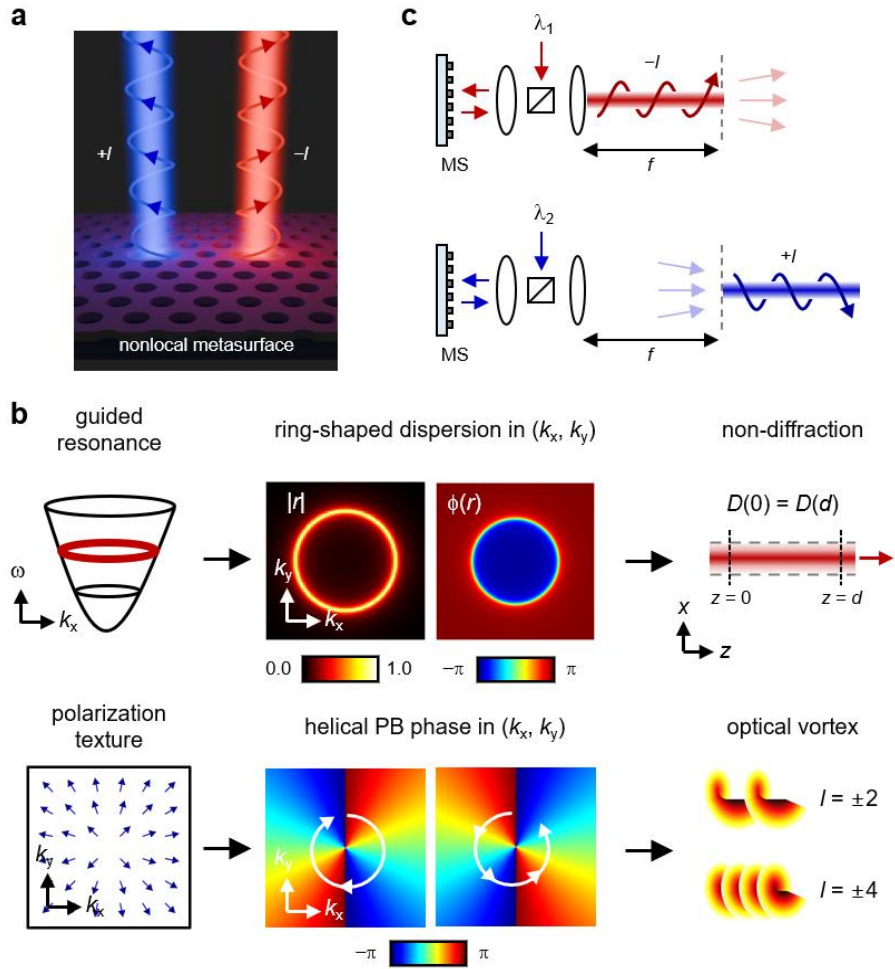

**Figure 1 Schematic of topological non-diffraction in a nonlocal metasurface. a,** Schematic of nonlocal metasurface with triangular lattice and generated optical vortex beams. **b,** Physical requirements for generating vortex Bessel beams, provided by the nonlocal metasurface. Guided resonances induce a ring-shaped dispersion in momentum space ($k_x$, $k_y$), enabling spatial non-diffraction. Simultaneously, the topological polarization texture gives rise to a helical Pancharatnam–Berry (PB) phase distribution, facilitating the generation of optical vortices. **c,** Schematic of vortex Bessel beam measurement in reflection. Depending on the operating wavelength ($\lambda_1$, $\lambda_2$), spatial asymmetry emerges in the propagation profile with respect to the location of metasurface in the imaging plane.

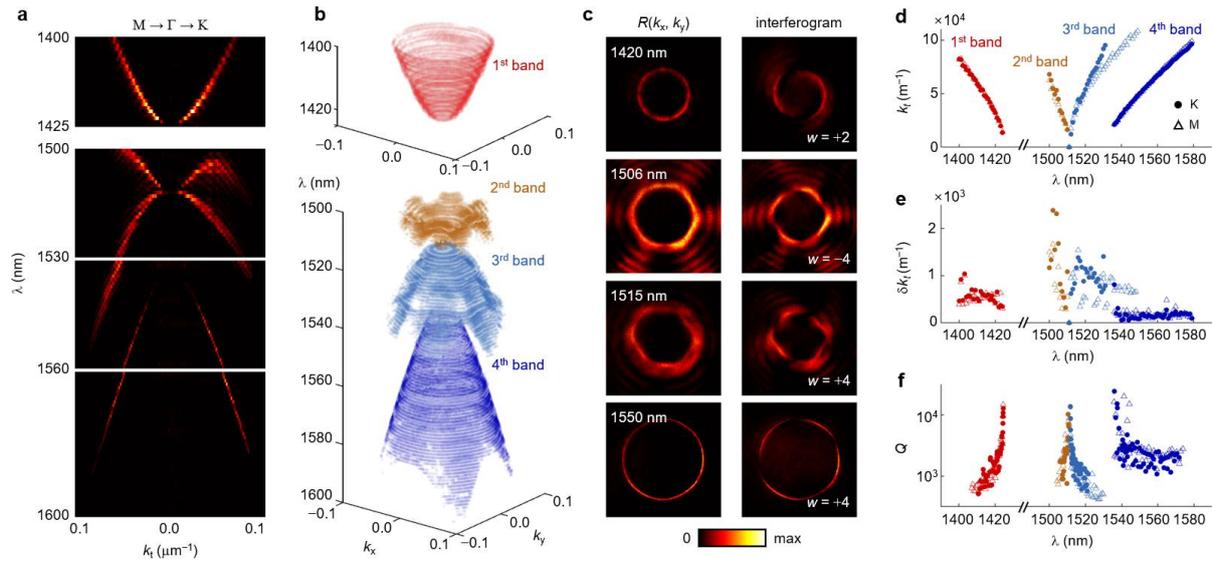

**Figure 2 Measurement of photonic bandstructure and topological characteristics. a,** Measured reflectance spectra of the nonlocal metasurface plotted in wavelength ($\lambda$) versus transverse wavevector ($k_t$) along the M–Γ–K trajectory in the first Brillouin zone. **b,** Three-dimensional visualization of the extracted photonic band structures: 1st (red), 2nd (orange), 3rd (sky blue), and 4th band (blue), plotted in ($k_x$, $k_y$, $\lambda$) space. **c,** Measured intensity distributions $R(k_x, k_y)$ (top row) and corresponding interferograms (bottom row) of reflected light in momentum space at representative wavelengths: 1420 nm, 1506 nm, 1515 nm, and 1550 nm. Spiral interference fringes confirm the presence of phase singularities with winding numbers $w = +2$, $-4$, $+4$, and $+4$, respectively. **d–f,** Extracted parameters for the four photonic bands: **d,** transverse wavevector magnitude ($k_t$); **e,** transverse wavevector linewidth ($\delta k_t$); **f,** quality factor $Q$. The narrow linewidth and high $Q$ values across all bands are consistent with the presence of symmetry-protected bound states in the continuum.

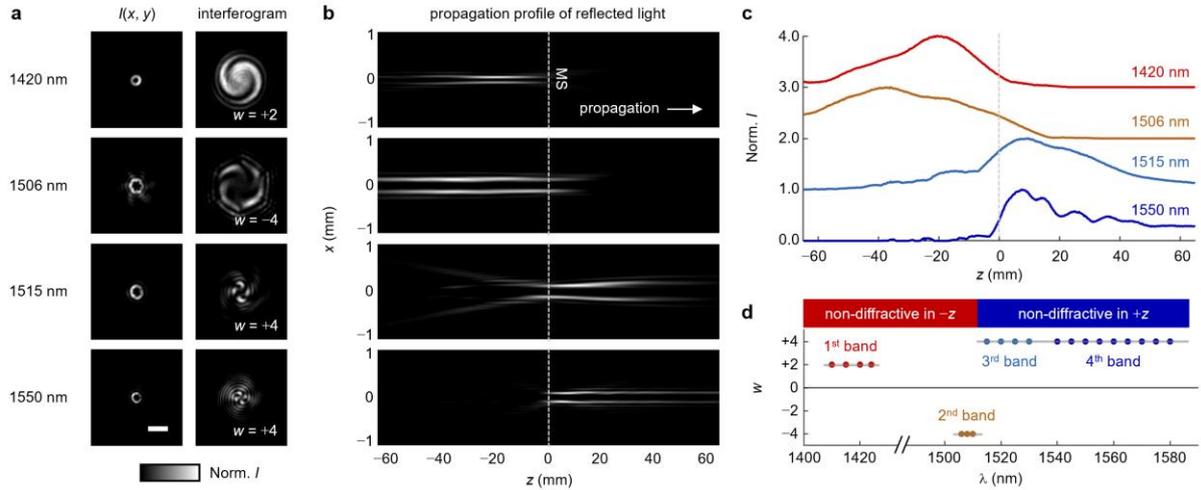

**Figure 3 Propagation profile of vortex Bessel beam. a,** Transverse intensity distributions $I(x, y)$ (top row) and corresponding interferograms (bottom row) of the reflected light in real space, measured at wavelengths of 1420 nm, 1506 nm, 1515 nm, and 1550 nm, respectively. The donut-shaped intensity profiles and spiral fringe patterns confirm the presence of optical vortices. Scale bar: 100 μm. **b,** Real-space propagation profiles of the reflected beams along the optical axis. The dashed line marks the location of the metasurface (MS). The reflected light is propagating toward +z-direction. **c,** Extracted longitudinal normalized intensity distributions $I(z)$ for each operating wavelength, highlighting the emergence of non-diffraction either on the +z or −z side depending on the photonic band. **d,** Summary of the observed asymmetric non-diffraction behavior (upper panel) and the corresponding optical vortex winding numbers (lower panel) as functions of operating wavelength.

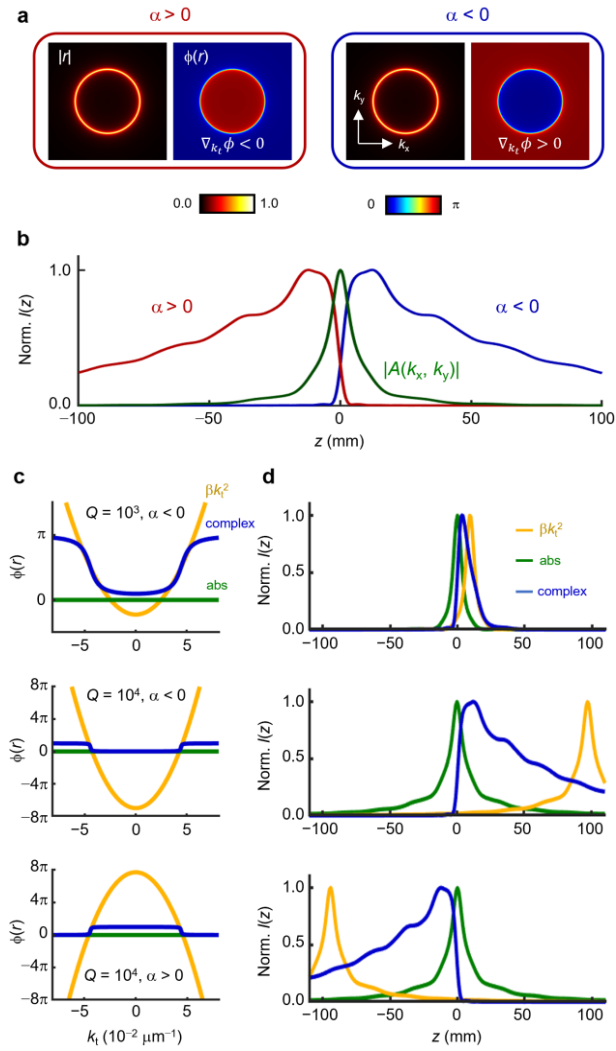

**Figure 4 Asymmetric non-diffraction governed by radial phase gradients in momentum space.
a,** Simulated amplitude $|r(k_x, k_y)|$ and phase $\phi_r(k_x, k_y)$ distributions of the reflection coefficient of guided resonance with positive ($\alpha > 0$, red box) and negative ($\alpha < 0$, blue box) band curvatures. **b,** Calculated longitudinal intensity distribution $I(z)$ of the reflected beam based on the full reflection coefficients in **a**. The red and blue curves correspond to $\alpha > 0$ and $\alpha < 0$, respectively. The green curve represents the case with a uniform phase (calculated using only $|r(k_x, k_y)|$). **c,** Phase profiles $\phi_r(k_t)$ (blue) extracted along radial $k_t$-axis for guided resonance with $Q = 10^3, 10^4$, fitted with a quadratic function $\beta k_t^2$ (yellow). The green curve represents the uniform-phase case. **d,** Calculated longitudinal intensity profiles $I(z)$ based on the phase distributions in **c**, using the same amplitude $|r(k_x, k_y)|$. The quadratic phase case (yellow) leads to beam displacement toward $z = -d_{\text{eff}}$, while full guided resonance phase (blue) yields an asymmetric propagation. The uniform-phase case (green) remains centered at the metasurface.

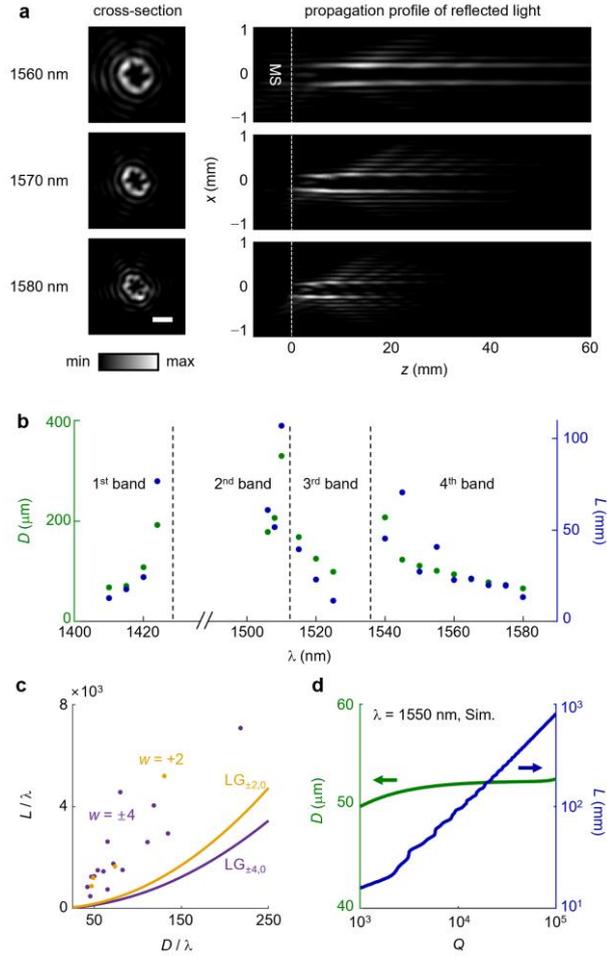

**Figure 5 Tunability of beam-diameter and propagation distance. a,** Transverse intensity distributions $I(x, y)$ (left) and real-space propagation profiles (right) of the reflected beams measured at wavelengths of 1560 nm, 1570 nm, and 1580 nm, respectively. Scale bar is 100 μm. The dashed line indicates the position of the metasurface (MS). The reflected light is propagating toward +$z$-direction. **b**, Measured beam diameter $D$ (green) and propagation distance $L$ (blue) of the vortex Bessel beams across operating wavelengths spanning the 1st to 4th photonic bands. **c,** Normalized beam diameter ($D/\lambda$) and propagation distance ($L/\lambda$) of metasurface-generated vortex beams with winding numbers $w = +2$ (yellow dot) and $w = \pm 4$ (purple dot), compared to the theoretical relation for conventional Laguerre-Gaussian (LG) beams: $LG_{\pm 2,0}$ (yellow line) and $LG_{\pm 4,0}$ (purple line). **d,** Numerically calculated beam diameter $D$ (green) and propagation distance $L$ (blue) of Bessel beams generated by guided resonance with the condition in Figure 4b depending on quality factor $Q$.